\documentclass{article}

\usepackage{PRIMEarxiv}

\usepackage[utf8]{inputenc} % allow utf-8 input
\usepackage[T1]{fontenc}    % use 8-bit T1 fonts
\usepackage{hyperref}       % hyperlinks
\usepackage{url}            % simple URL typesetting
\usepackage{booktabs}       % professional-quality tables
\usepackage{amsfonts}       % blackboard math symbols
\usepackage{amsmath}
\usepackage{nicefrac}       % compact symbols for 1/2, etc.
\usepackage{microtype}      % microtypography
\usepackage{lipsum}
\usepackage{fancyhdr}       % header
\usepackage{graphicx}       % graphics
\usepackage{natbib}
\usepackage{eurosym}
\usepackage{booktabs,tabularx}
\usepackage{rotating}
\usepackage{pdflscape}
\usepackage{graphicx}
\usepackage{subcaption}

\title{Simulating Ride-Pooling Services with Pre-Booking and On-Demand Customers
}

\author{
  Roman Engelhardt\\
  Chair of Traffic Engineering and Control \\
  Technical University of Munich \\
  Arcisstraße 21 \\
  80333 Munich, Germany \\
  \texttt{Corresponding Author: roman.engelhardt@tum.de} \\
  \And
  Florian Dandl \\
  Chair of Traffic Engineering and Control \\
  Technical University of Munich \\
  Arcisstraße 21 \\
  80333 Munich, Germany \\
  \texttt{florian.dandl@tum.de} \\
  \And
  Klaus Bogenberger \\
  Chair of Traffic Engineering and Control \\
  Technical University of Munich \\
  Arcisstraße 21 \\
  80333 Munich, Germany \\
  \texttt{klaus.bogenberger@tum.de} \\
}

\begin{document}
\maketitle

\begin{abstract}
If private vehicle trips can be replaced, ride-pooling services can decrease parking space needed by higher vehicle utilization and increase traffic efficiency by increasing vehicle occupancy. Nevertheless, substantial benefits can only be achieved if a certain market penetration is passed to find enough shareable rides for pooling to take place. Additionally, because of their highly dynamic and stochastic nature on-demand ride-pooling services cannot always guarantee that a request is served. Allowing customers to pre-book their trip in advance could provide benefits for both aspects. Additional knowledge helps an operator to better plan vehicle schedules to improve service efficiency while an accepted trip or a rejection can be communicated early on to the customer. This study presents a simulation framework where a ride-pooling provider offers a service in mixed operation: Customers can either use the service on-demand or pre-book trips. A graph-based batch optimization formulation is proposed to create offline schedules for pre-booking customers. Using two rolling horizons, this offline solution is forwarded to an online optimization for on-demand and pre-booking customers simultaneously. The framework is tested in a case study for Manhattan, NYC. That the graph-based batch optimization is superior to a basic insertion method in terms of solution quality and run-time. Due to additional knowledge, the ride-pooling operator can improve the solution quality significantly by serving more customers while pooling efficiency can be increased. Additionally, customers have shorter waiting and detour times the more customers book a trip in advance.
\end{abstract}

% keywords can be removed

\section{Introduction}
Growth in population and urbanization stresses the capacity of urban transportation infrastructure all over the world. Since space is limited, more efficient ways to utilize the existing infrastructure are required. With the rise of on-demand mobility, also on-demand ride-pooling (ODRP) systems emerged where customers --- after requesting a trip on demand --- are matched together and share the space on the road with travelers going into a similar direction. In contrast to ride-sourcing services like Uber or Didi, in this study the operator of the ODRP service centrally dispatches a fleet of vehicles to serve these customers while trying to pool multiple customers into the same vehicle to share a route. Especially with the introduction of autonomous vehicles and low operating costs in sight~\cite{Bosch.2018}, these services might significantly reduce the number of vehicles within a city and increase average vehicle occupancy if private vehicle trips can be replaced.

To quantify the performance of ODRP systems, several simulation studies have been conducted recently for different cities and different demand settings. On the one hand, studies assuming a large market penetration for ride pooling systems (e.g.~\cite{AlonsoMora.2017} or \cite{Fiedler.2018}) show huge benefits regarding fleet sizing and traffic impacts when replacing taxi trips in New York City or private vehicle trips in Prague, respectively. On the other hand, studies for Munich \cite{Engelhardt.2019} or Austin \cite{Fagnant.2018} assuming lower market penetrations of ODRP services showed that up to 50k customers per day are needed for pooled miles traveled to overcome empty miles traveled. These results, along with macroscopic models \cite{Tachet.2017, Bilali.2020}, reveal the strong scaling property of ODRP-Systems; substantial benefits for a traffic system can only be achieved if a certain threshold in market penetration is passed. This threshold is especially critical in stages of early implementation and less dense environments.

A possibility to improve service performance is to allow pre-bookings for customers. This study investigates a mixed ride-pooling system, where customers have the options to pre-book or request a ride on demand. Despite losing flexibility, pre-booking can have advantages for customers as well as operators: Because of the dynamic and stochastic nature of the ride-pooling problem, it cannot be guaranteed that all customers can be served within certain time constraints. Pre-booking a trip can remove these insecurities because a guaranteed pick-up or rejection can be communicated a long time in advance. From a customer's point of view, ride-pooling might be an convenient option for many trips that can be planned ahead of time and do not need to be on demand, e.g., a trip to work the next day or to the cinema at the weekend. From an operator’s standpoint, pre-bookings can reduce dynamism and stochasticity of the underlying vehicle routing problem. With additional knowledge of revealed future trips compared to an on-demand-only system, schedules can be planned early on, promising an improvement in the overall operator objectives~\cite{Bilali.2019b}. However, with more knowledge and a larger solution space to optimize, the computational complexity increases. Moreover, pre-bookings are no guarantee for operational efficiency in a mixed system with pre-booked and on-demand customers, as also downsides for an operator are imaginable: The constraints of already confirmed and guaranteed reserved trips might prohibit the assignment of potentially more on-demand customers, who would be favorable with respect to the operator’s objectives.

\section{Literature Review}

The vehicle routing problem for ride-pooling services can be defined as a variant of the dial-a-ride-problem (DARP). The goal of this optimization problem is to compute optimal vehicle routes (i.e. by minimizing cost) while satisfying a set of constraints. Usually the pick-up is constraint by a time-window and the drop-off is constraint by a maximum travel time. A first dynamic programming approach has been proposed by \citep{Psaraftis.1980}; since then the methodology has been extended, e.g., by \citep{Cordeau.2006}. Nevertheless, due to the NP-hardness of the problem the problem size was limited to four vehicles with up to 30 customers to be transported in the later. To increase the feasible problem size, heuristics (i.e. \cite{Jaw.1986}) and meta-heuristics (i.e.~\cite{Cordeau.2003, Jain.2011, Parragh.2013, Massobrio.2016}) have been proposed. Nevertheless, the scale remained rather small as, for example, an instance with a maximum of 45 passengers has been solved in~\cite{Massobrio.2016}. For recent developments, the reader is referred to \citep{Molenbruch.2017}, who provided a review for solution approaches for the DARP.

The problem of controlling the ride-pooling services as considered in this study differs from the DARPs being solved in the mentioned studies above: Firstly, especially on-demand request create a highly dynamic and stochastic system in contrast to often used DARP instances that mostly consider a-priori knowledge of demand and secondly, employing a city-wide service requires matching hundreds to thousands of vehicles to possibly hundreds of thousands customers a day in real-time. Further criteria for differentiation can be found in \cite{Hyland.2017}. While the dynamic and stochastic component is mostly removed for pre-booking customers, the system size make the previously mentioned solution approaches still not applicable for finding solutions for the assignment of pre-booked customers.

The approach for solving the assignment problem for on-demand ride-pooling services usually consists of solving the static variant for currently active customer requests (i.e. customers that just requested a trip, wait for the pick-up or are already on-board of a vehicle). Heuristic~\cite{Jaw.1986} and meta-heuristic~\cite{Jung.2016} approaches have been proposed for finding solutions in real-time. On the other hand, the graph-based approach by \citep{AlonsoMora.2017} based on shareability graphs~\cite{Santi.2014} allows finding any-time optimal solutions for large scale applications. Further heuristics based on this algorithm allow speed-up with only minor solution deterioration~\cite{Simonetto.2019, Engelhardt.29.07.2020}. To increase fleet efficiency (i.e. reduced vehicle kilometers (VKT) or increased vehicle occupancy) algorithms that dynamically select pick-up and drop-off locations~\cite{Fielbaum.2021, Engelhardt.2021} or allow transfers between vehicles~\cite{Masoud.2017} have been proposed.

The short-coming of only using information about currently revealed customers is that information about future requests is ignored and vehicles can end up in areas with low demand while they are missing in areas with high demand. To balance demand and supply distributions, re-positioning algorithms can be applied. \citep{Zhang.2016} proposed an algorithm that repositions idle vehicles according to a solution of a minimum transportation problem matching expected future demand and supply per zone. \citep{Syed.2021} extended the formulation to incorporate the spatial spread of zones. Former algorithms are explicitly formulated for the ride-hailing use case where a vehicle can only transport one customer at a time. For the ride-pooling use case a reactive repositioning algorithm based on sending vehicles to positions where customers could not be served was proposed~\cite{AlonsoMora.2017} while a pro-active algorithm samples possible future requests from a forecast distribution and inserts them in the assignment problem~\cite{AlonsoMora.2017.2}.

Similarly to the last approach, pre-booked trips allow pre-positioning of vehicles according to these request, which in contrast to~\cite{AlonsoMora.2017.2} are known with certainty. To exploit the complete knowledge of the pre-booked trips, the approach in this study is to first create schedules for the fleet's vehicles to serve the pre-booked trips as good as possible and adopt this solution online to also serve incoming on-demand customers. Therefore, algorithms are required that can solve large scale instances of the static assignment problem. \citep{Su.2020} proposed a meta-heuristic to find initial solutions for purely pre-booked ride-pooling services. The algorithm proposed by \citep{Kucharski.2020} compares the utility of pooled with hailed rides to create utility based shareability graphs that can be used to create solutions for static variant. This study adopts a graph-based method originally designed to compute the necessary fleet size and initial vehicle distribution to serve a certain demand \cite{Wallar.52020195242019}.

To the authors knowledge, only two other studies consider the mixed mobility on-demand service with on-demand and pre-booking customers: \citep{Duan.2020} propose an assignment algorithm for the ride-hailing use-case. \citep{Abkarian.2022} evaluates a ride-pooling services, but instead of booking trips ahead, customers reserve vehicles completely for a certain time duration for private use.

\section{Contribution}

In this study a graph-based algorithm for a ride-pooling service is proposed that allows assigning both on-demand and pre-booked customers. Customers pre-book the service and their request is confirmed (or rejected) a day ahead. This confirmation is considered binding for the operator. The solution quality of initial schedules created for the pre-booked trips is compared to an insertion heuristic. Additionally, the impact of the quality of this static initial solution onto the efficiency of the dynamic mixed operation is analyzed. The benefit for fleet operators of allowing pre-booking is elaborated on a case study for Manhattan, NYC.

\section{Methodology}

This methodology section is structured as follows: First, the problem is described. Secondly, the general simulation framework and the simulation flow is introduced. In a third step, the online optimization algorithm is defined. Fourthly, the methodology to create initial solution for pre-booked trips is introduced. And finally, the integration of on-demand trips and pre-booked trips is described.

\subsection{Problem Description}

%\begin{figure}[hb]
%     \centering
%     \begin{subfigure}[b]{0.49\textwidth}
%         \includegraphics[width=\textwidth]{figs_new/Example_1a.png}
%         \label{fig:ex1a}
%     \end{subfigure}
%     \hfill
%     \begin{subfigure}[b]{0.49\textwidth}
%         \includegraphics[width=\textwidth]{figs_new/Example_1b.png}
%         \label{fig:ex1b}
%     \end{subfigure}
%     \hfill
%     \begin{subfigure}[b]{0.49\textwidth}
%         \includegraphics[width=\textwidth]{figs_new/Example_2a.png}
%         \label{fig:ex2a}
%     \end{subfigure}
%     \hfill
%     \begin{subfigure}[b]{0.49\textwidth}
%         \includegraphics[width=\textwidth]{figs_new/Example_2b.png}
%         \label{fig:ex2b}
%    \end{subfigure}
%     \hfill
%     \begin{subfigure}[b]{0.6\textwidth}
%         \includegraphics[width=\textwidth]{figs_new/example_legend.png}
%         \label{fig:exleg}
%     \end{subfigure}
%        \caption{Two Examples sketching the possible impact of pre-bookings on the ODRP-Systems. In Example 1 the number of served customers can be increased. In Example 2 less customers are served because of pre-booking.}
%        \label{fig:example_sketch}
%\end{figure}

In this study, two types of customers request trips from a ride-pooling provider: 1) On-demand customers that request an as-soon-as-possible service. 2) Pre-booking customers that request a trip at a specific pick-up time in the future. Both types of customers expect an answer whether they can be served shortly after sending their request. The operator's answer is supposed to be binding, which is especially relevant for pre-booked trips as the fleet control algorithm has to ensure that those customers are scheduled even though the state of the fleet and the demand at the time of the pre-booked trip is unknown. The operator controls a fleet of vehicles providing the service. The goal of the control algorithm is to assign schedules, i.e. an ordered list of customer pick-up and drop-offs, to its vehicles that ensure time and capacity constraints while maximizing a certain objective (e.g. revenue). With respect to a pure on-demand service, mainly three aspects change: 1) The operator has access to additional information, because a part of the customer requests are known ahead. 2) An answer to a customers whether a requested trip is accepted or rejected should be provided not long after the request. 3) The binding answer adds another constraint to the vehicle routing problem.

\subsection{General Simulation Framework}

\begin{figure}[!ht]
  \centering
  \resizebox{0.8\textwidth}{!}{\includegraphics{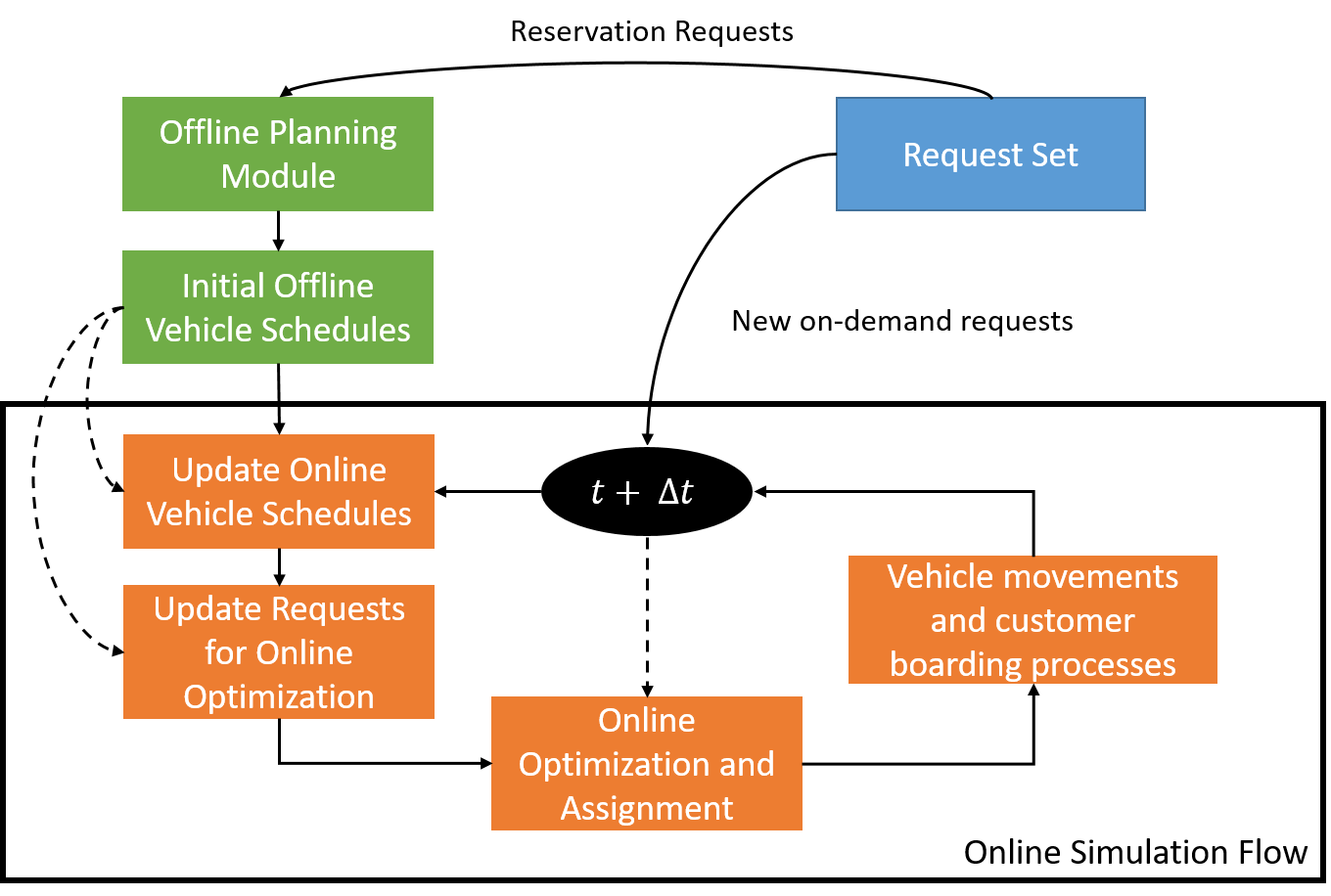}}
  \caption{High-level flowchart of the simulation framework.}\label{fig:flowchart}
\end{figure}

The simulations are performed within the open-source fleet simulation tool FleetPy\footnote{https://github.com/TUM-VT/FleetPy}~\cite{Engelhardt.28.07.2022}, which is extended to fit the needs of this study. FleetPy is an agent-based simulation framework to study mobility-on-demand services. The agents used in this study are (i) customers requesting a trip (on-demand or pre-booked) from an (ii) operator offering a ride-pooling service by assigning routes to its (iii) vehicles that serve the requested trips. The high-level simulation framework is sketched in Figure \ref{fig:flowchart}.

Vehicles move in a network $G = (N,E)$ with nodes $N$ and edges $E$. Each edge $e \in E$ is associated with a travel time and a travel distance. A customer request $r_i$ is described by the tuple $(o_i, d_i, t_i, t_i^{e})$, with $o_i \in N$ and $d_i \in N$ being origin and destination location of the requested trip, respectively. $t_i$ represents the time of the request. $t_i^{e}$ corresponds to the earliest pick-up time. In this study it is assumed that all pre-booked trips are known beforehand and therefore the set of requests $R$ can be split in on-demand requests $R_o$ and pre-booked requests $R_p$. For on-demand requests $r_i \in R_o$, $t_i = t_i^{e}$ applies. For pre-booked requests $r_i \in R_p$, $t_i = 0$s is assumed. 

Before the simulation starts, the operator computes initial plans for pre-booked trips in the offline planing module, described in the next section. These initial schedules are available to the operator at the beginning of the simulation and are used within the online optimization to serve on-demand and pre-booked trips simultaneously.

During each time step of $\Delta t$, new on-demand customers request trips from the fleet operator. The online optimization module computes new schedules for the the fleet's vehicles in batch and replies to customers if they can be served within certain time constraints. If a customer cannot be served, it leaves the system unserved. In this study it is assumed that schedules have to fulfill certain constraints regarding customer pick-up and drop-off to be feasible: 
\begin{enumerate}
    \item No pick-up is allowed before the earliest pick-up time $t_i^e$.
    \item No pick-up is allowed after the latest pick-up time $t_i^l = t_i^e + \delta_{max}^{wait}$, with a maximum waiting time $\delta_{max}^{wait}$.
    \item The maximum in-vehicle travel time $tt_i^{max}$ may no exceed $tt_i^{direct}(1 + \delta_{max}^{detour})$. $tt_i^{direct}$ is the shortest possible travel time $tt_i^{direct}$ for customer $i$ to drive from origin to destination and $\delta_{max}^{detour}$ a detour factor.
    \item The number of customers on-board of a vehicle my not exceed its capacity $c_v$.
\end{enumerate}
This constraints have to be fulfilled for on-demand and pre-booking customers. If no schedule can be found to satisfy all constraints, a request is declined.

The fleet operator rates feasible schedules $\psi_k(v;R_\gamma)$ by an objective function $\phi$ with the goal to assign schedules that minimize the overall objective. Hereby, the schedule $\psi_k(v;R_\gamma)$ refers to the $k$-th feasible permutation of boarding stops for vehicle $v$ serving the set of requests $R_\gamma$. The objective function in this study is defined by
\begin{equation}
\label{eq:obj}
    \phi(\psi_k(v;R_\gamma)) = c_{dis} \cdot d(\psi_k(v;R_\gamma)) + c_{vot} \cdot \left(\sum_{i\in R_\gamma} t_i^{do}(\psi_k(v;R_\gamma)) - t_i^{e}\right) - P \cdot |R_\gamma| .
\end{equation}
$c_{dis}$ and $c_{vot}$ are cost coefficients weighting driven distance and customer trip duration, respectively. $d(\psi_k(v;R_\gamma))$ refers to the distance vehicle $v$ has to drive to perform this schedule. $t_i^{do}(\psi_k(v;R_\gamma))$ expresses the drop-off time of customer $i$. $|R_\gamma|$ refers to the number of customers being served by the schedule. $P$ is a large reward parameter to prioritize assigning as many customers as possible.

At the end of the simulation time step, vehicles move in the network according to their plans and scheduled boarding and alighting processes for customers are performed.

\subsection{Creation of Pre-Booking Schedules}
Finding the optimal solution of multiple thousands of reservation requests as required in this study is computationally intractable. Therefore, heuristic methods are needed to create initial pre-booked schedules for vehicle. In this section, two heuristic methods are described to create initial vehicle schedules for pre-booked trips.

\subsubsection{Insertion (I) Method}
As a baseline, the first method is based on an insertion heuristic. Requests are inserted successively into the current schedules. If insertions generate feasible schedules for a request, the schedule minimizing the difference in objective function of Eq.~\ref{eq:obj} is accepted. Because multiple thousand of pre-booking requests have to be computed, this insertion heuristic method can take a lot of computational time if all current vehicle plans are checked. Therefore, the search space is additionally decreased by only trying to insert a reservation request $r_i$ into the current schedules of $N_{insert}^{max, v}$ vehicles based on their expected proximity to $o_i$ at time $t_i^e$. For vehicles without currently assigned schedule, its initial position is used at the expected position. For vehicles with assigned schedule, the last scheduled stop before $t_i^e$ is used as its expected position. Despite its simplicity, the disadvantage of the I-Method is that usually very long schedules have to be checked for feasibility after insertion; especially when pre-bookings are considered a long time ahead. 

\subsubsection{Forward Batch Optimization (FBO) Method}
The proposed FBO-Method avoids the previously described problem by grouping reservation requests into batches firstly. In a second step, all feasible schedules within each batch are computed. Lastly, batches are connected and schedules within the batches are assigned by solving an Integer Linear Program (ILP). This connection is designed to maintain feasibility of schedules created within the batches, avoiding the need for feasibility checks of very long schedules like in the I-Method.

To this end, reservation requests are sorted by their earliest pick-up time. A maximum of $N_{batch}^{max, s}$ adjacent requests are then grouped into a batch. Therefore, the probability of finding shareable trips but also computational effort within these batches increases with $N_{batch}^{max, s}$. The batches can be sorted based on earliest pick-up times of containing requests.

The creation of all feasible schedules within a batch is based on the algorithm proposed by \citep{AlonsoMora.2017} and exploits the formulation of shareability networks introduced by \citep{Santi.2014}. Because these schedules should be defined independently of vehicles, the vehicle independent schedule $\psi_k(-;R_\gamma)$ is defined as a scheduled list of boarding stops which could be performed by a hypothetical vehicle immediately available at the first boarding stop. Additionally, a request bundle $\Xi(-;R_\gamma)$ is defined as the collection of all feasible schedules $\psi_k(-;R_\gamma)$ that serve the same set of requests. The objective of a request bundle is defined by
\begin{equation}
    \label{eq:rb-obj}
    \phi(\Xi(-;R_\gamma)) = \min_k \phi(\psi_k(-;R_\gamma)) .
\end{equation}
The grade of a request bundle is defined by the number of requests served by this bundle. The creation of all feasible schedules can now be executed iteratively: Each request is consecutively inserted into the stack of existing request bundles thereby incrementing their grade. Starting with bundles of grade 1, the procedure forms a new request bundle with grade increased by one if a feasible schedule can be found. To increase efficiency of the algorithm the following check, which is necessary for the existence of a request bundle of higher grade, can be made: A request bundle $\phi(\Xi(-;R_\gamma))$ of grade $N$ can only exist if all request bundles $\phi(\Xi(-;R_\gamma \backslash r_l)) \forall r_l \in R_\gamma$ of grade $N-1$ exist. Hence, one can only find a feasible schedule for the set of requests $(r_1, r_2, r_3)$ if feasible schedules for $(r_1, r_2)$, $(r_2, r_3)$ and $(r_1, r_3)$ exist.

Once all feasible schedules within a batch are computed, schedules from one batch have to be connected with schedules from other batches to form executable schedules for the whole simulation horizon. This optimization problem is defined on a directed graph $G_{opt} = (N_{opt}, E_{opt})$. A node $m$ on this graph corresponds to a request bundle and is represented by the tuple $(\phi_m, o_m, d_m, t_m^{start}, t_m^{end})$, with the objective value $\phi_m$, the first boarding location $o_m$, the last boarding location $d_m$, the earliest start time at the first boarding location $t_m^{start}$ and the planned end time at the last boarding location $t_m^{end}$. Additional to the request bundle, nodes for vehicles are added to guarantee feasible assignments. For a vehicle node the tuple reads $(0, x_m, x_m, t_m^0, t_m^0)$ with $x_m$ being the initial vehicle position and $t_m^0$ the earliest time the vehicle is available (simulation start time at beginning of the simulation). \\
Starting from an empty set of edges, an edge $(m, n)$ from node $m$ to node $n$ is added to $E_{opt}$ if
\begin{equation}
\label{eq:sched_constr}
    t_n^{start} \leq t_m^{end} + tt(d_m, o_n),
\end{equation}
with $tt(d_m, o_n)$ the (expected) travel time from $d_m$ to $o_n$. This inequality ensures that all constraints within the schedule of $n$ can still be fulfilled, if $n$ is scheduled after $m$. Each edge is associated with the cost $c_{m, n} = c_{dis} \cdot d(d_m, o_n) + \phi_n$. Because Eq. \ref{eq:sched_constr} ensures that the planned times for boarding stops are not delayed if request bundles are scheduled, this cost is in line with Eq.~\ref{eq:obj}. Lastly, each node is connected to a hypothetical node $\nu$ with cost $0$ and this node is connected to all vehicle nodes to enable a flow-conserving constraint in the formulation of the optimization problem.

The optimization problem to schedule request bundles and assign them to vehicles can now be formulated as follows:
\begin{align}
    \label{eq:batchopt1}
    \text{minimize} \qquad & \qquad \sum_{ (m,n) \in E_{opt} } c_{m,n} \cdot z_{m,n} & \\
    \label{eq:batchopt2}
    \text{s.t.} \qquad & \qquad \sum_{m, n} z_{m,k} - z_{k,n}  = 0  &\forall k \in N_{opt} \\
    \label{eq:batchopt3}
        & \qquad \sum_m z_{m,n}  \leq 1  &\forall n \in N_{opt} \setminus \{\nu\} \\
    \label{eq:batchopt4}
        & \qquad \sum_n z_{m,n}  \leq 1  &\forall m \in N_{opt} \setminus \{\nu\} \\
    \label{eq:batchopt6}
        & \qquad \sum_m z_{m,\nu}  \leq |V|  & \\
    \label{eq:batchopt7}
        & \qquad \sum_n z_{\nu,n}  \leq |V|  & \\
    \label{eq:batchopt8}
        & \qquad \sum_m \sum_{n \in \Omega_l} z_{m,n}  \leq 1  &\forall r_l \in R 
\end{align}
The goal of the optimization problem is to assign edges $(m,n)$ that minimize the overall cost (usually $c_{m,n} \leq 0$ holds because of the assignment reward in Eq.~\ref{eq:obj}). The decision variable $z_{m,n} \in \{0,1\}$ therefore takes the value $1$ in Eq.~\ref{eq:batchopt1} if nodes $m$ and $n$ are assigned and scheduled. The constraint in Eq.~\ref{eq:batchopt2} describes the flow conservation constraint where each assigned node must have the same number of incoming and an outgoing assigned edges. Eq.~\ref{eq:batchopt3} and Eq.~\ref{eq:batchopt4} constrain that each node may only be assigned once. Eq.~\ref{eq:batchopt6} and Eq.~\ref{eq:batchopt7} constraint the hypothetical node $\nu$ to the number of vehicles. The formulation of Eq.~\ref{eq:batchopt8} ensures that each request is only assigned at most once. To this end, $\Omega_l$ refers to the set of nodes (request bundles) containing request $r_l$. 

The number of variables of this optimization problem becomes to high to solve large scale instances. On the one hand, the number of feasible schedules (nodes) scales exponentially with $N_{batch}^{max, s}$. On the other hand, connections between batches (edges) with a high temporal separation are nearly always feasible. Therefore, the problem is decomposed into smaller sub-problems that are solved successively: Let $L$ be the sorted list of batches based on the included earliest pick-up times of requests. Instead of solving the optimization problem of Eq.~\ref{eq:batchopt1} for the whole list of sorted batches $L[j]$ $(j \in \mathbb{N}_0)$, only the sub-problem of the set $[L[j], L[j + 1], ..., L[j + N_{batch}^{c}]$ is solved. Thereby, one starts with $j=0$. The initial vehicle distribution is used as start constraints. After the solution is obtained, the assignment for the first batch in the list is fixed and determines the new vehicle constraints, when solving the problem for the set $[L[j + 1], L[j + 2], ..., L[j + N_{batch}^{c}+1]$. This procedure continues until the last batch of the sorted batch list.

\begin{figure}[!ht]
  \centering
  \resizebox{0.95\textwidth}{!}{\includegraphics{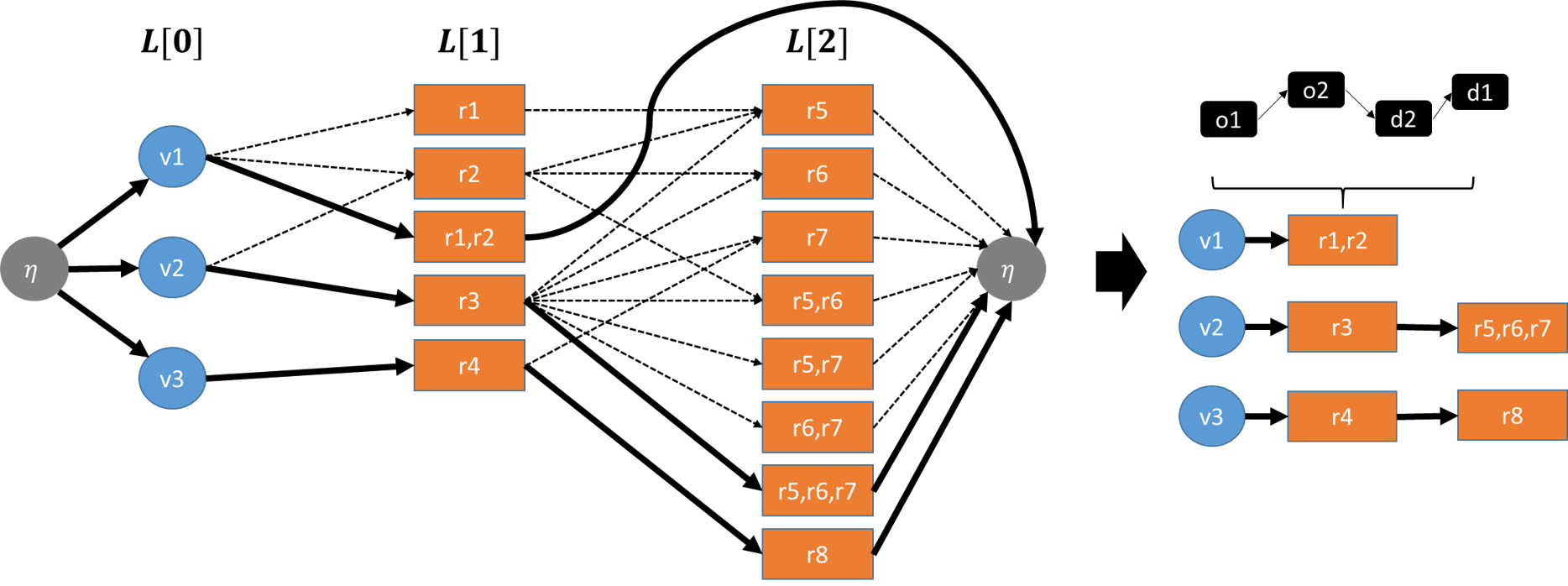}}
  \caption{Sketch for creating schedule for pre-booking requests. For clarity only connection between neighboring batches are shown (there are also connection between e.g. $L[0]$ and $L[2]$). Additionally, incoming and outgoing connections of node $\nu$ refer to a single hypothetical node. Bold connection correspond to assignments after solving the ILP.}\label{fig:fbo}
\end{figure}

The FBO algorithm is therefore described by the two hyper-parameters $N_{batch}^{max, s}$ (the number of requests per batch) and $N_{batch}^{c}$ (the number of batches taken into account for the assignment problem). Shared routes can only be created within the batches, $N_{batch}^{max, s}$ therefore controls the computational effort spent to create these routes. $N_{batch}^{c}$ on the other hand controls the time span the scheduling optimization process is able to look into the future. Hence, the scheduling process can be interpreted as a rolling horizon approach. A sketch of the whole procedure is shown in Fig.~\ref{fig:fbo}.

\subsection{Integrating Pre-Booking Schedules into Online Optimization}
To integrate pre-booking and an on-demand requests for the ride-pooling service, the dispatching algorithm incorporates on-demand requests into the pre-computed schedules for pre-booked requests. A naive approach to solve this problem would be to assign the offline schedules at the beginning of the simulation and insert on-demand requests into this solution when requested. Nevertheless, this approach does not allow for dynamic global optimization of the current fleet state (i.e. no reassignments of pre-booked requests would be possible). This is especially relevant when the fraction on on-demand requests is much higher than pre-booked requests. In this setting, it can no longer be assumed that the solution for pre-booked requests is good enough compared to the optimal solution of all revealed requests. Additionally, the approach is computationally not efficient because the whole offline schedules are always considered even if planned stops far in the future are not directly relevant for the assignment of on-demand requests.

The approach in this study is to introduce two different rolling horizons: The first is the short-term horizon $T_h^{short}$. Once the horizon $t_s + T_h^{short}$ with the simulation time $t_s$ approaches the earliest pick-up time of a reservation request, the request is added to the set of of online requests and will now be considered in the online optimization. The second horizon is the revelation horizon $T_h^{rev} \geq T_h^{short}$. Requests with earliest pick-up times after this horizon are not revealed to the online optimization explicitly. Only the first stop following this horizon is revealed to the online optimization acting as an end constraint of the schedule: No stop is allowed to be inserted after this stop. This constraint ensures that the requests that are assigned after this horizon in the offline solution remain feasible during the online optimization. In a nutshell, the short-term horizon constrains the problem size of the online optimization while reassignments are still possible, while the revelation horizon constrains the length of schedules that have to be checked for feasibility and feasibility of pre-booked requests in the future.

\begin{figure}[!ht]
  \centering
  \resizebox{0.8\textwidth}{!}{\includegraphics{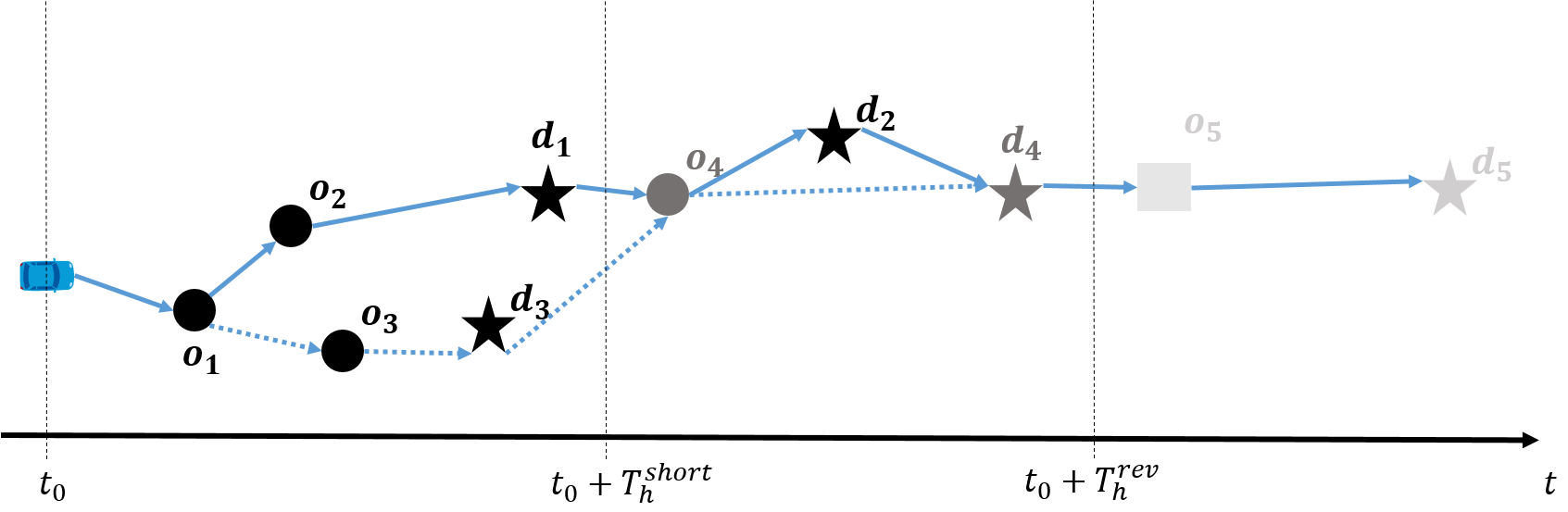}}
  \caption{Sketch showing the effect of different rolling horizons. Customers with earliest pick-up time before $t_0 + T_h^{short}$ are marked in black. Theses customers can be re-assigned within the online optimization indicated by an alternative dashed route. Customers with earliest pick-up time between $t_0 + T_h^{short}$ and $t_0 + T_h^{rev}$ are already revealed to the online optimization but cannot be re-assigned to other vehicles. The online optimizer is not able to see stops after the first stop ahead of $t_0 + T_h^{rev}$ marked with the light grey rectangle. The algorithm is forbidden to alter the stops after the horizon. Hence, the approach does not consider the drop-off of request $5$ explicitly in the example.}\label{fig:horizons}
\end{figure}

\subsection{Online Optimization}
For the online optimization an algorithm inspired by~\cite{AlonsoMora.2017} is used. While a high level description is presented in the following, the reader is referred to~\cite{Engelhardt.29.07.2020} for details in implementation.

The idea of the algorithm is to first build all feasible schedules for all requests considered in the online optimization and then solve an ILP for assigning the optimal ones to vehicles. While an exhaustive search is computationally intractable, a guided search analogously to the creation of schedules for pre-booked requests can be used. Let $\psi_{k}(v;R_\gamma)$ be the $k$-th feasible permutation of stops for vehicle $v$ serving the set of requests $R_\gamma$. Similar to the request bundle, the vehicle-to-request-bundle (V2RB) $\Xi(v;R_\gamma)$ is defined as the collection of feasible permutations $k$ for vehicle $v$ serving the set of requests $R_\gamma$. Again, the objective value of the V2RB is defined by
\begin{equation}
    \phi(\Xi(v;R_\gamma)) = \min_k \phi(\psi_{k}(v;R_\gamma)) ,
\end{equation}
with $\phi$ defined by Eq.~\ref{eq:obj}.

The following three necessary conditions for the possible existence of a V2RB can be used for a guided search process:
\begin{enumerate}
    \item a V2RB $\Xi(v;{r_l})$ serving a single request $r_l$ can only exist, if the vehicle $v$ can reach $o_l$ before the latest pick-up time elapsed.
    \item any V2RB serving the two requests ${r_l, r_m}$ can only exist if the request bundle $\Xi(-,{r_l, r_m})$ is feasible.
    \item similar to request bundles, a V2RB $\Xi(v;R_\gamma)$ of grade $N$ can only exist, if all V2RBs of grade $N-1$ are feasible when removing any of the requests in $R_\gamma$.
\end{enumerate}
Following these conditions, all feasible V2RBs can be created by inserting requests into corresponding V2RBs of lower grade.

With respect to incorporating reservation schedules, two minor adjustments are made. Firstly, only V2RBs for requests with earliest pick-up times within $T_h^{short}$ are created. Secondly, to ensure that assigned pre-booked schedules remain feasible, feasible insertions always have to fulfill the condition that the start of a connected pre-booked schedule (the grey rectangle in Fig.~\ref{fig:horizons}) have to be reachable in time.

Finally, the assignment of online schedules is made by solving the following ILP:
\begin{align}
    \label{eq:onlineopt1}
    \text{minimize} \qquad & \qquad \sum_v \sum_\gamma \phi(\Xi(v;R_\gamma)) \cdot z_{v,\gamma} & \\
    \label{eq:onlineopt2}
    \text{s.t.} \qquad & \qquad \sum_\gamma z_{v,\gamma}  \leq 1 \qquad &\forall v \in V_o \\
    \label{eq:onlineopt3}
        & \qquad \sum_v \sum_{\gamma \in \Omega_i} z_{v,\gamma}  = 1 \qquad &\forall i \in R_a \\
    \label{eq:onlineopt4}
        & \qquad \sum_v \sum_{\gamma \in \Omega_i} z_{v,\gamma}  \leq 1 \qquad &\forall i \in R_u
\end{align}
The decision variable $z_{v,\gamma} \in {0,1}$ in Eq.~\ref{eq:onlineopt1} takes the value $1$ if the V2RB $\Xi(v;R_\gamma)$ is assigned. Constraint \ref{eq:onlineopt2} ensures that at most one V2RB is assigned for each vehicle. Constraint \ref{eq:onlineopt3} ensures that each already assigned request is assigned again (but reassignments to other vehicles are possible). $\Omega_i$ corresponds to the set of V2RBs the customer $i$ can be served with. Constraint \ref{eq:onlineopt4} ensures that not yet assigned customers are assigned at most once.

\subsection{Repositioning}
For repositioning idle vehicles into regions with less supply than demand, the reactive reposition algorithm described in \cite{AlonsoMora.2017} is applied. Each time an on-demand request could not be served, its origin is marked as repositioning target. Only currently idle vehicles are considered; moreover, if a pre-booked assignment is scheduled the vehicles are supposed have enough time to serve demand in the destination area before serving the next pre-booked assignment. Hence, to be considered for a repositioning to position $p_r$, a vehicle $v$ with pre-booked assignment is only considered if the start time of the next stop is further then $T_{repo} + tt(p_v, p_r) + tt(p_r, p_b)$ ahead, with $tt(x, y)$ as the travel time between position $x$ and $y$ and $p_v$ the position of the vehicle, $p_r$ the position of the repositioning target and $p_b$ the origin position of the next pre-booked customer. The goal of the repositioning assignment is to assign idle vehicles to these repositioning targets while the overall travel time is minimized. For vehicles with a pre-booked assignment, the effective travel time to the repositioning target is considered to be $tt_{eff}(p_v, p_r) = tt(p_v, p_r) + tt(p_r, p_b) - tt(p_v, p_b)$. Thereby trips are prioritized that reposition vehicles closer to position of the next pre-booked trip.

\section{Case Study}

The proposed methods are tested for a case study for Manhattan, New York City using the publicly available taxi data set\footnote{https://www1.nyc.gov/site/tlc/about/tlc-trip-record-data.page}.

The network $G=(N,E)$ has been extracted from OpenStreetMap using the Python library OSMnx~\cite{Boeing.2017}. Instead of using free flow travel times, the method described by \citep{Syed.2021} is used to compute scale factors for edge travel times to resemble the travel times provided in the taxi data set. Based on resulting hourly scaling factors, a daily average for the day 2018/11/12 is used in this study. Taxi trips from 2018/11/12 are used as requests for the ride-pooling service. From $202569$ overall trips, the data set is sub-sampled to $10$\% of the overall trips by randomly removing $9$ out of $10$ trips. $3$ input files with different random seeds of the sub-sampling process are created and used for each simulation presented. 

Requests are randomly split into on-demand and pre-booking customers. Shares of $0$\%, $10$\%, $25$\%, $50$\%, $75$\% and $100$\% pre-booking customers are tested in the simulations, with $25$\% corresponding to the base case for an analysis of the hyper-parameter sensitivities. The fleet size $|V|$ is determined by running calibration simulations without pre-booking customers. A fleet size of $|V| = 200$ vehicles is picked that serves approximately $90\%$ of the on-demand customers.

Table \ref{tab:parameters} summarizes all input parameters and their base values for the simulation. Those base values are used in the simulations if not explicitly stated otherwise.

\begin{table}[]
\begin{tabular}{l|l|l}
Parameter               & Description                                                                                 & Base Value \\ \hline
$\Delta t$              & Simulation Time Step                                                                        & $60$s      \\
$\delta_{max}^{wait}$   & Maximum Customer Waiting Time                                                               & $6$min     \\
$\delta_{max}^{detour}$ & Maximum Customer Detour Factor                                                              & $40$\%     \\
$c_{dis}$               & Distance Dependent Cost Factor                                                              & $0.25$€/km~\cite{Bosch.2018} \\
$c_{VOT}$               & Value of Time                                                                               & $16.2$€/h~\cite{Frei.2017}  \\
$P$                     & Assignment Reward                                                                           & $10000$€   \\
$N_{batch}^c$           & Number of Batches Within optimization (FBO)           & $2$        \\
$N_{batch}^{max, s}$    & Maximum Number of Requests per Batch (FBO)                    & $20$       \\
$N_{insert}^{max, v}$   & Maximum Number of Vehicles Searched (I)                                      & $25$       \\
$T_h^{short}$           & Short-Term Horizon                             & $720$s     \\
$T_h^{rev}$             & Revelation Horizon                          & $720$s    \\
$T_{repo}$              & Minimum Time Until Next Pre-Booked Stop for Repositioning & $3600$s \\
$|V|$       & Number of Vehicles (Fleet Size) & $200$ \\
$c_v$  & Capacity of Vehicles & $4$
\end{tabular}
\caption{Main parameters of the simulation framework. Values in the result section correspond to the given base value as long as not explicitly mentioned.}
\label{tab:parameters}
\end{table}

\section{Results}

In the following, simulation results of the case study are presented. First, the two algorithms to create schedules for pre-booking customers are compared followed by a sensitivity analysis for the two introduced rolling horizons. At the end the share of pre-booking customers is varied to evaluate the impact on the operation and customer experience. All computations are implemented in Python and conducted on an Intel Xeon Silver processor with $2.10$GHz and up to $192$GB RAM. Optimization problems are solved with the commercial solver "Gurobi"\footnote{https://www.gurobi.com/}.

\subsection{Creation of Pre-booking Schedules}

\begin{figure}
     \centering
     \begin{subfigure}[b]{0.49\textwidth}
         \includegraphics[width=\textwidth]{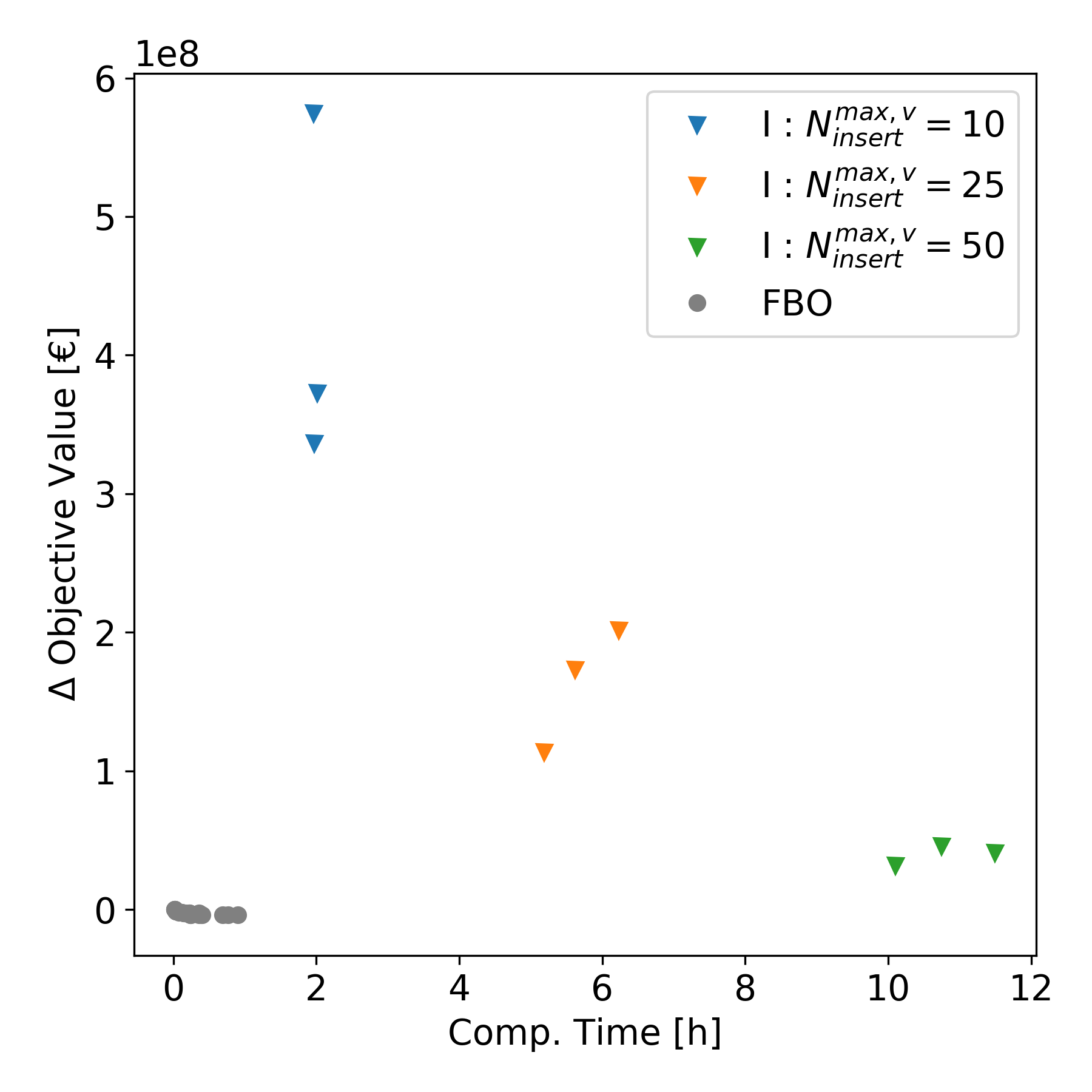}
         \caption{Comparison FBO- and I-Algorithm.}
         \label{fig:off_sol_a}
     \end{subfigure}
     \hfill
     \begin{subfigure}[b]{0.49\textwidth}
         \includegraphics[width=\textwidth]{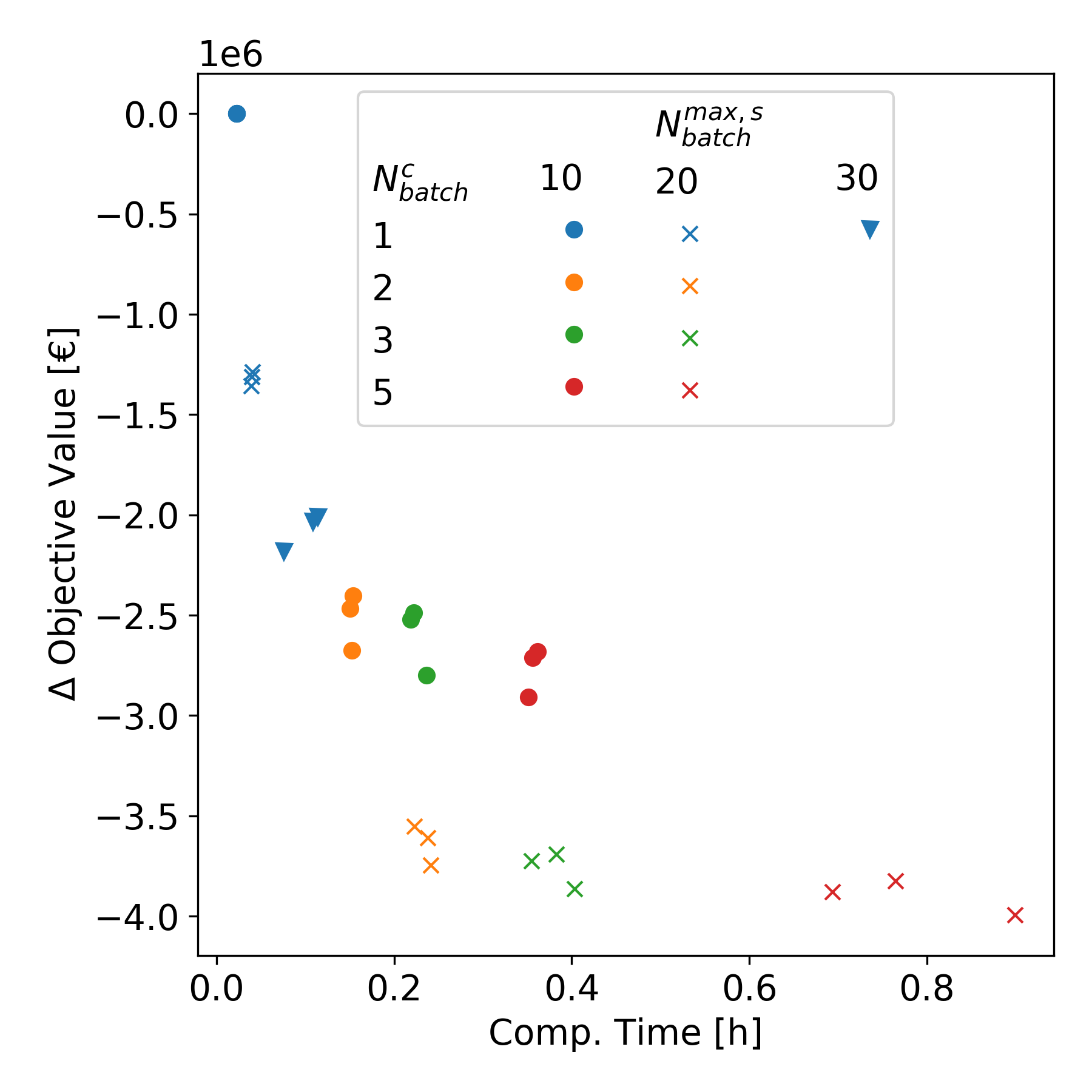}
         \caption{Only FBO-Algorithm}
         \label{fig:off_sol_b}
    \end{subfigure}
        \caption{Comparison of computational time and solution quality for different parameter sets of the two proposed algorithms to compute initial solutions for the reservation requests. All objective values shown are relative to the objective values of the FBO-method scenario with parameters $(N_{batch}^{max, s} = 10, N_{batch}^{c} = 1)$ of the corresponding request set.}
        \label{fig:off_sol}
\end{figure}

Fig.~\ref{fig:off_sol} shows the comparison of the two presented algorithms to create the schedules for pre-booking customers for different parameters of the algorithms. For each parameter set, three different seeds (request sets) are simulated and illustrated. Each plot shows the computational time of the algorithm against the overall objective value which has to be minimized. The objective values are evaluated relative to the reference case with the same request set and the FBO-Method with $N_{batch}^{c} = 1$ and $N_{batch}^{max, s} = 10$.

Fig.~\ref{fig:off_sol_a} compares the FBO- and the I-Method. All parameter sets for the FBO-Method are shown as grey dots. It can be seen that for all scenarios tested the FBO-Method is superior to the I-Method: A lower objective value is generated with much lower computational time. The I-method can produce better results with higher $N_{insert}^{max, v}$ because more vehicles are searched for possible insertions into the current schedule. Nevertheless, because in this schedule always the whole schedule has to be checked for feasibility resulting in high computational times. By splitting the requests into batches, the FBO-method only has to check shorter temporaly constraint schedules after a new insertion.

Fig.~\ref{fig:off_sol_b} illustrates results of the FBO-Method only (grey dots in Fig.~\ref{fig:off_sol_a}). Colors represent the  number of requests per batch $N_{batch}^{max, s}$ while same markers indicate the same number of consecutive batches $N_{batch}^{c}$ considered in the assignment problem. The solution quality can be improved with increasing $N_{batch}^{max, s}$ and $N_{batch}^{c}$. Increasing $N_{batch}^{c}$ from 1 to 2 is especially beneficial for finding better solutions with only a minor increase in computational time while a further increase does not improve the overall objective value significantly. It should be noted that due to the high number of possible routes within the batch with $N_{batch}^{max, s} = 30$ and the resulting large number of variables the optimization problem, only a problem variant with $N_{batch}^{c} = 1$ could be solved optimally.

Because of a good trade-off between solution quality and computational time, all following computations are carried out with the FBO-method and the parameters $N_{batch}^{max, s} = 20$ and $N_{batch}^{c} = 2$.

\subsection{Integrating Pre-Booking Schedules into Online Optimization}

\begin{figure}
     \centering
     \begin{subfigure}[b]{0.49\textwidth}
         \includegraphics[width=\textwidth]{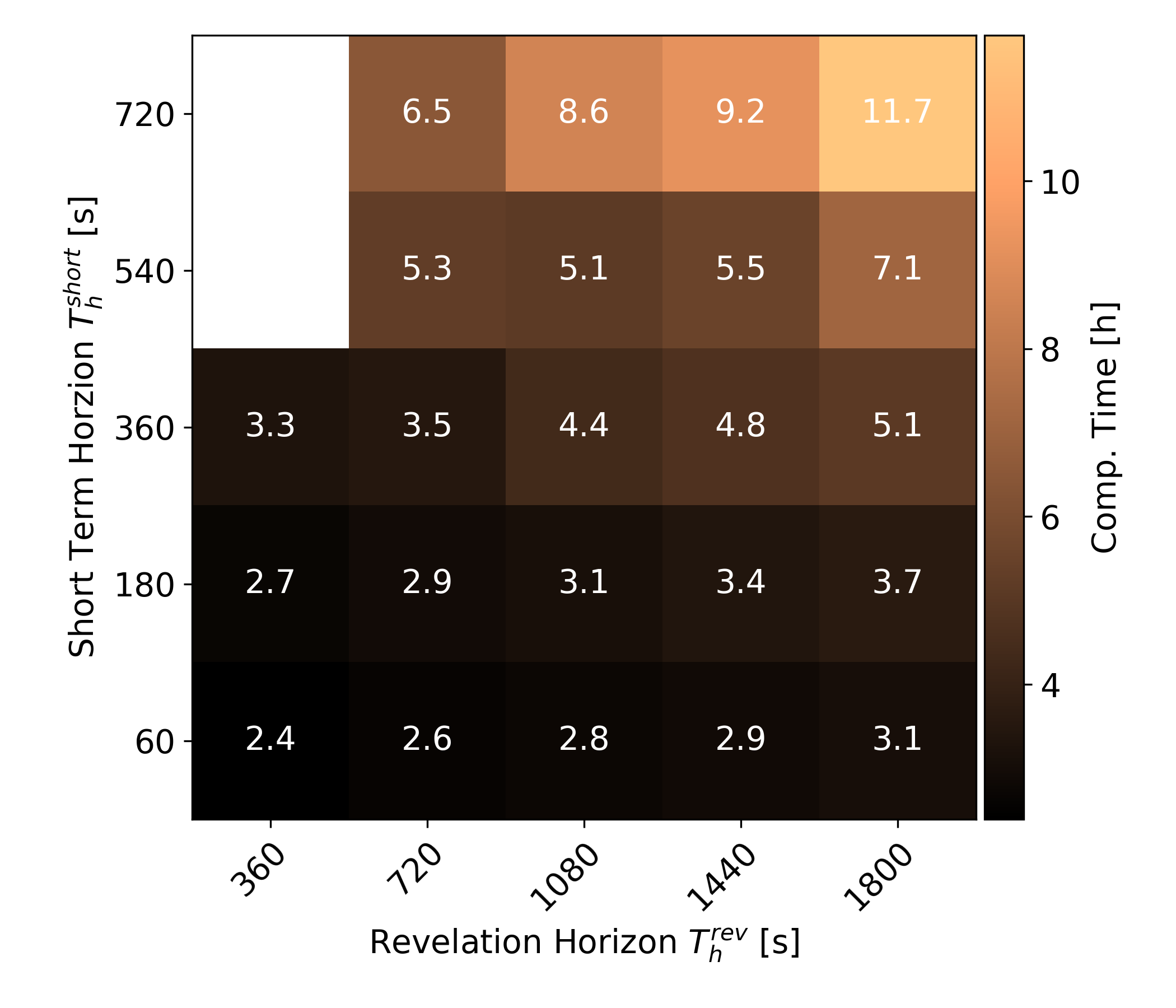}
         \caption{Computational Time}
         \label{fig:on_off_a}
     \end{subfigure}
     \hfill
     \begin{subfigure}[b]{0.49\textwidth}
         \includegraphics[width=\textwidth]{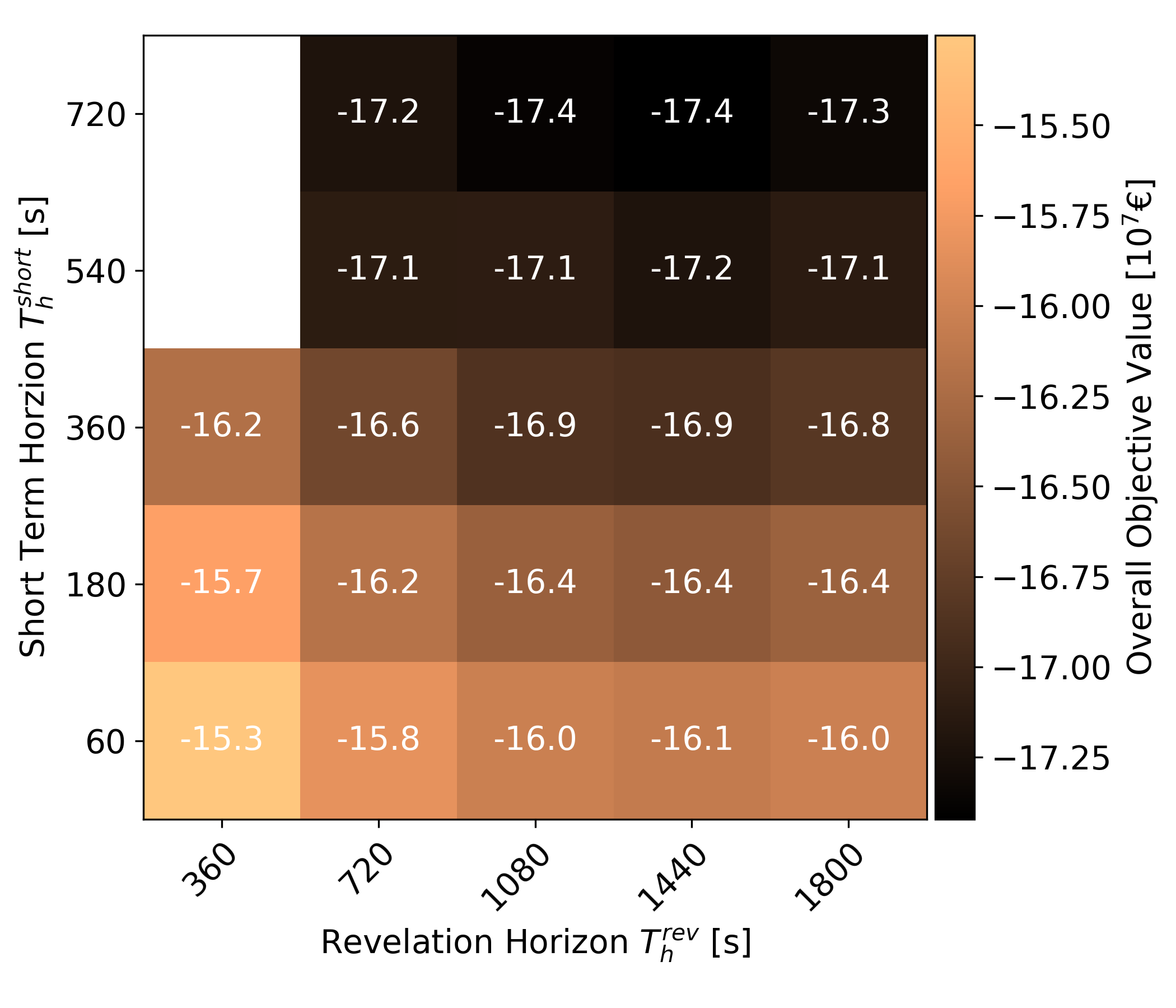}
         \caption{Overall Objective Value}
         \label{fig:on_off_b}
    \end{subfigure}
        \caption{Computational time and online solution objective for varying parameters of $T_h^{rev}$ and $T_h^{short}$.}
        \label{fig:matrix_on_off}
\end{figure}

In a next step the sensitivity of the horizon parameters $T_h^{rev}$ and $T_h^{short}$ for inserting the offline solution for pre-booked trips into the online algorithm are compared. Fig.~\ref{fig:matrix_on_off} compares the impact of various combinations for $T_h^{rev}$ and $T_h^{short}$ on the overall computational time and objective value to be minimized. White spaces indicate undefined parameter combinations ($T_h^{rev} < T_h^{short}$).

Generally, it can be observed that with higher $T_h^{rev}$ and $T_h^{short}$ the overall objective value improves while computational time increases. With longer $T_h^{short}$ pre-booked requests can be reassigned by the online optimization earlier. Therefore, the online optimization can adopt the vehicle routes of pre-booked customers earlier according to on-demand requests. As trade-off, computational time increases due to a growing solution space. With longer $T_h^{rel}$ additional pre-booked stops are revealed to the online optimization, which cannot be assigned to other vehicles but new on-demand request can be inserted to fit future stops. Therefore, the objective decreases with higher $T_h^{rel}$ but computational time increases because longer schedules have to be checked for feasibility. A strong decrease in the overall objective (and increase in computational time) can be observed between the step from $T_h^{short} = 360$s to $T_h^{short} = 540$s. For higher values of $T_h^{short}$ the improvement of solution quality is minor. This effect might result from the waiting time constraint $\delta_{max}^{wait} = 360$s of on-demand requests. If $T_h^{short} > \delta_{max}^{wait}$, the algorithm has sufficient look-ahead to reschedule pre-booked requests and accommodate on-demand requests better.

Because of a good trade-off between computational time and solution quality, the parameter set $T_h^{short} = T_h^{rel} = 720$s is used for further simulations.

\subsection{Impacts on Ride-Pooling Service}

\begin{figure}
     \centering
     \begin{subfigure}[b]{0.49\textwidth}
         \includegraphics[width=\textwidth]{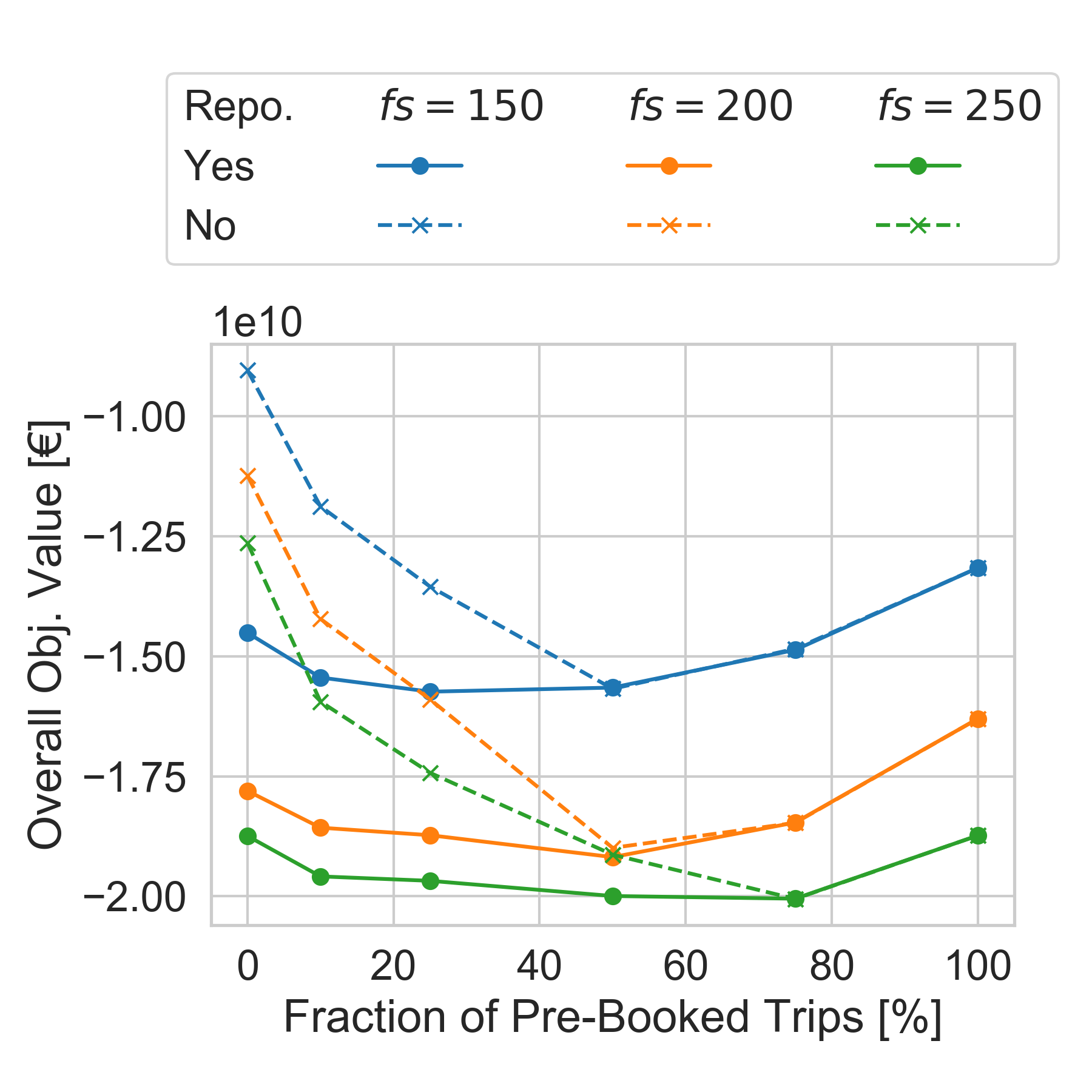}
         \caption{}
         \label{fig:dem_obj}
     \end{subfigure}
     \hfill
     \begin{subfigure}[b]{0.49\textwidth}
         \includegraphics[width=\textwidth]{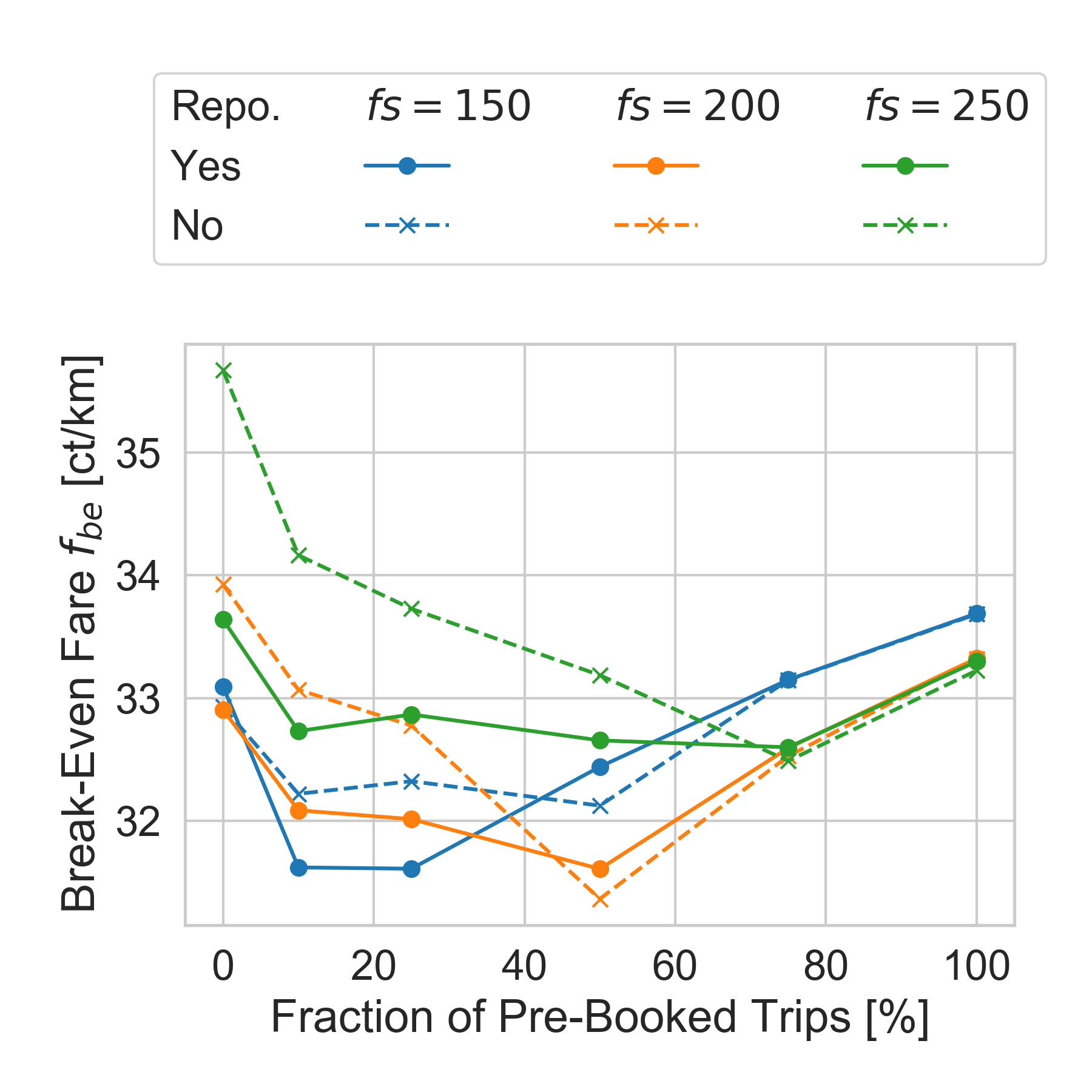}
         \caption{}
         \label{fig:dem_break_even}
     \end{subfigure}
        \caption{Overall Objective and Break Even Fares for different penetrations of pre-booking customers.}
        \label{fig:dem_op_1}
\end{figure}

Finally, different key performance indicators (KPIs) quantifying the impacts on the ride-pooling service are quantified. In Fig.~\ref{fig:dem_op_1} the overall objective value and the break-even fare are compared for different penetration rates of pre-booked trips, for different fleet sizes and with and without repositioning of idle vehicles. Hereby, the distance-dependent break-even fare $f_{be}$ is computed by solving

\begin{equation}
    C_v \cdot |V| + c_d \cdot d_{fleet} = f_{be} \sum_{i \in R_{served}} d_i^{direct} ~.
\end{equation}

\begin{figure}
     \centering
     \begin{subfigure}[b]{0.40\textwidth}
         \includegraphics[width=\textwidth]{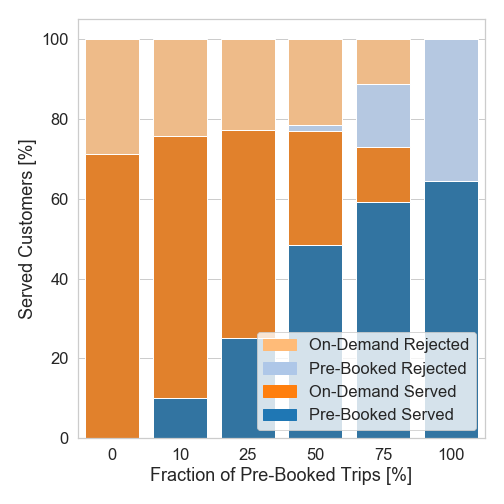}
         \caption{Fleetsize: 150, With Repo.}
         \label{fig:served1b}
    \end{subfigure}
     \hfill
     \begin{subfigure}[b]{0.40\textwidth}
         \includegraphics[width=\textwidth]{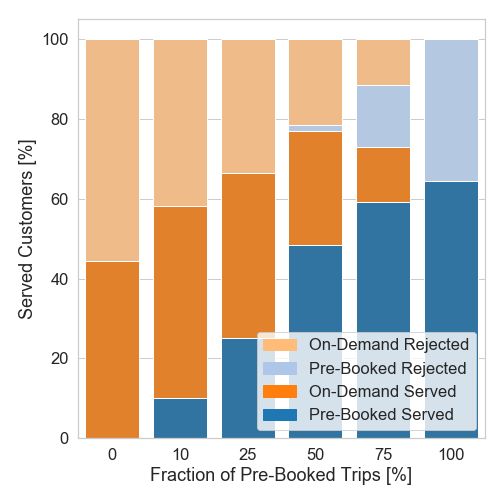}
         \caption{Fleetsize: 150, No Repo.}
         \label{fig:served1a}
     \end{subfigure}
    \hfill
     \begin{subfigure}[b]{0.40\textwidth}
         \includegraphics[width=\textwidth]{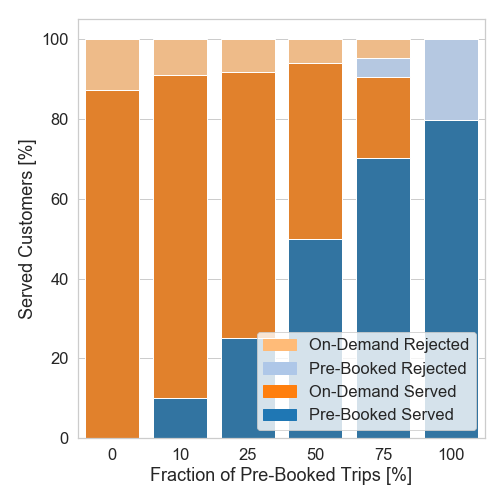}
         \caption{Fleetsize: 200, With Repo.}
         \label{fig:served1d}
    \end{subfigure}
     \hfill
     \begin{subfigure}[b]{0.40\textwidth}
         \includegraphics[width=\textwidth]{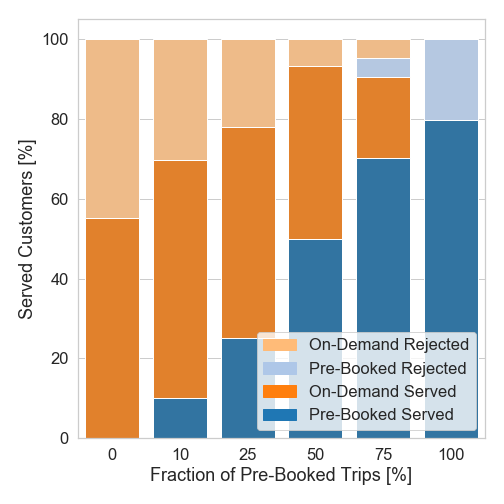}
         \caption{Fleetsize: 200, No Repo.}
         \label{fig:served1c}
     \end{subfigure}
    \hfill
     \begin{subfigure}[b]{0.40\textwidth}
         \includegraphics[width=\textwidth]{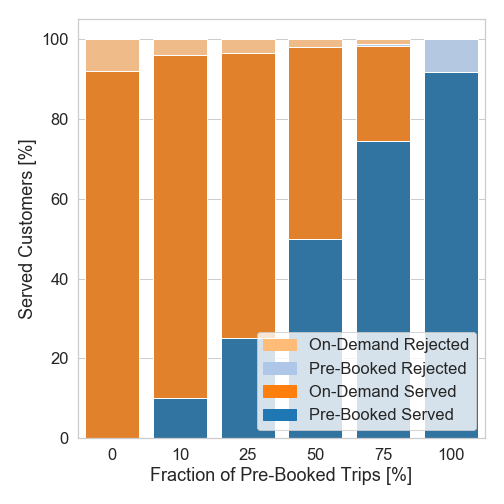}
         \caption{Fleetsize: 250, With Repo.}
         \label{fig:served1f}
     \end{subfigure}
    \hfill
     \begin{subfigure}[b]{0.40\textwidth}
         \includegraphics[width=\textwidth]{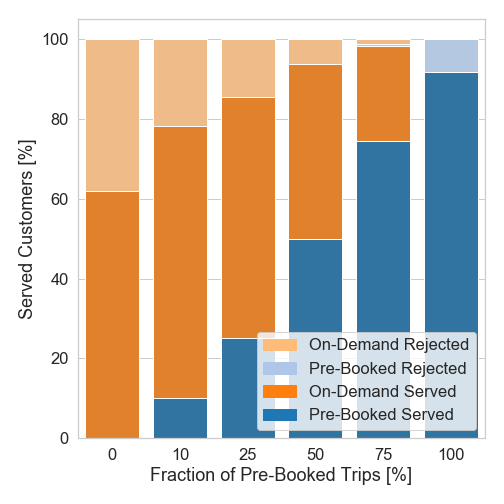}
         \caption{Fleetsize: 250, No Repo.}
         \label{fig:served1e}
    \end{subfigure}
        \caption{Served on-demand and pre-booked trips for different penetrations of pre-bookings and fleet sizes.}
        \label{fig:dem_served}
\end{figure}

$C_v = 25$€ is the fix cost for a vehicle for one day of operation and $c_d = 0.25$€/km is the distance dependent cost while $d_{fleet}$ is the driven distance of the fleet~ \cite{Bosch.2018,Dandl.2019b}. $d_i^{direct}$ is the direct route distance of a customer and $R_{served}$ is the set of served customers. Fig.~\ref{fig:dem_op_1} shows that in all settings the overall objective value decreases until a fraction of 50\% pre-booked trips, indicating that the ride-pooling service benefits from pre-bookings due to additional knowledge. This decrease is smaller when a repositioning algorithm is activated, indicating that positioning of vehicles based on expected demand already provides huge benefits to the system. Additionally, it can be observed that no scenario with pre-bookings and without repositioning has a lower overall objective value than the corresponding scenario with repositioning and without pre-booking. This indicates that pre-booking is not able to replace repositioning. For fractions of pre-booking trips above 50\% the overall objective starts to deteriorate. This mainly results from a bad solution quality of the schedules from pre-booked trips: When the demand of pre-booked trips reaches the capacity of the fleet, the offline algorithm is not able to accommodate all pre-booking customers (which can also be observed in Fig.~\ref{fig:dem_served}). This, in turn, leads to a degrading of the objective value due to the high weight on served customers. This effect can likely be avoided by an improved algorithm to create pre-booked schedules and is not an effect of pre-booking itself. 

Similar trends are observable for the break-even fare illustrated in Fig.~\ref{fig:dem_break_even} as it can be reduced until 50\% penetration of pre-booking customers. However, large fleet sizes are penalized due to fixed costs for this performance indicator. For a fleet size of $200$ vehicles the break-even fare can be reduced from $33.9$ct/km without repositioning to $31.4$ct/km ($7.3$\%) with 50\% pre-booking trips. The difference of $2.5$ct/km can be interpreted as additional revenue of the operator per customer-km or as fare reduction the operator can provide to customers when enabling repositioning and pre-booking.

Fig.~\ref{fig:dem_served} shows distributions of served and rejected pre-booking and on-demand customers. For low fractions of pre-booking customers, only on-demand customers are rejected because all pre-booked customers can be accommodated and the accepted reservation is binding to the operator. In all scenarios, pre-booking increases the overall number of served customers until pre-booked customers are rejected, because the offline algorithm no longer finds suitable routes to serve them all. Repositioning significantly increases vehicle availability and therefore the number of served requests. High fractions of pre-booking trips also leads to higher number of served on-demand customers indicating a better vehicle distribution.

\begin{figure}
     \centering
     \begin{subfigure}[b]{0.49\textwidth}
         \includegraphics[width=\textwidth]{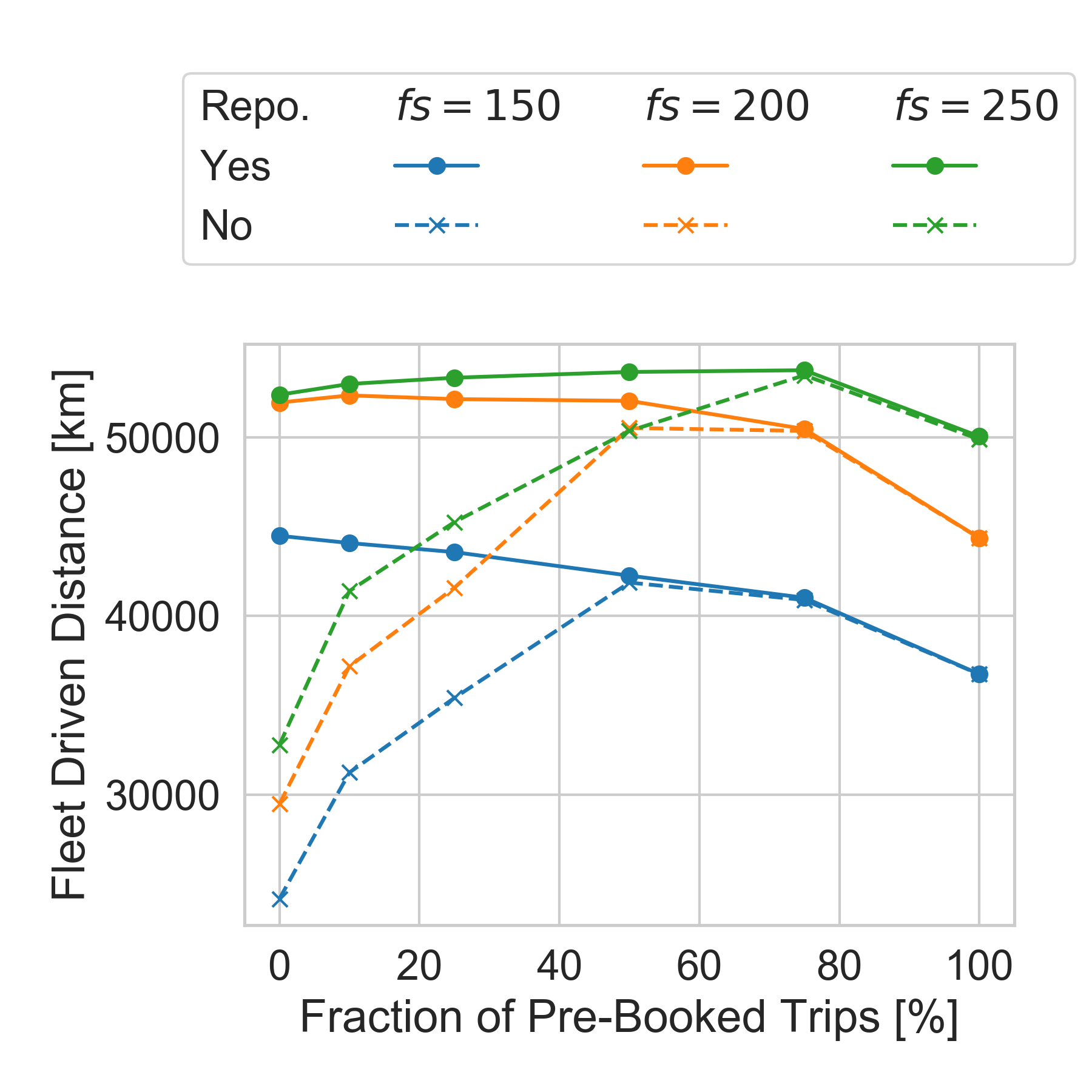}
         \caption{}
         \label{fig:vkta}
     \end{subfigure}
     \hfill
     \begin{subfigure}[b]{0.49\textwidth}
         \includegraphics[width=\textwidth]{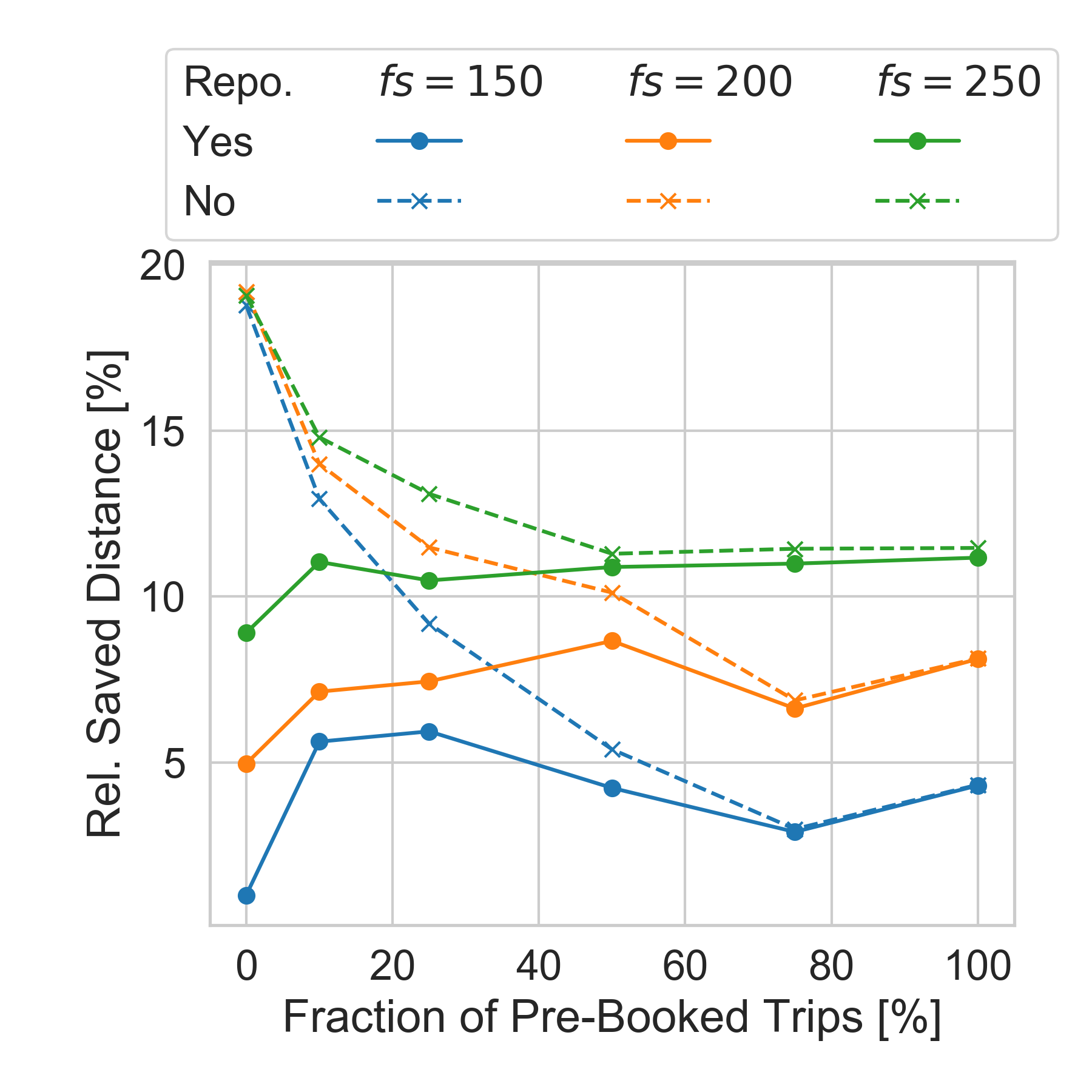}
         \caption{}
         \label{fig:savedb}
     \end{subfigure}
        \caption{Fleet driven distance and relative saved distance for different penetrations of pre-booking customers}
        \label{fig:dem_op_2}
\end{figure}

KPIs regarding fleet kilometers are illustrated in Fig.~\ref{fig:dem_op_2}. Without repositioning, the fleet kilometers are increasing with the number of pre-booked trips as the overall number of served customers increases. Due to higher fleet utilization and more served customers, the driven distance of the fleet is higher with repositioning; when also pre-booking is enabled, fleet kilometers increase only marginally (and even decrease for a fleet size of $150$ vehicles) while the number of served customers increases until at least 50\% pre-booked trips, indicating that rides can be shared efficiently. 

To reduce the dependency on the number of served requests, the KPI "Relative Saved Distance" is defined by 
\begin{equation}
  rsd = \frac{\sum_{i \in R_{served}} d_i^{direct} - d_{fleet}}{\sum_{i \in R_{served}} d_i^{direct}} ~.
 \end{equation}
This KPI, which is illustrated in Fig.~\ref{fig:savedb}, measures the sharing efficiency by comparing fleet driven distance with the distance customers would drive when traveling the direct distance. With repositioning this KPI rises when increasing the share of pre-booking trips from 0\% to 10\% and remains rather constant for higher fractions for a fleet size of $200$ and $250$ vehicles. In all scenarios with repositioning the relative saved distance is higher with pre-booking than without pre-booking indicating a higher pooling efficiency. Without repositioning and pre-booking, vehicles are restricted to pick-up new customers in the vicinity of their current position, which limits the number of served requests severely. Introducing pre-booking allows these requests to be served as well, but they often incur longer empty pick-up trips, thereby leading to a lower value in relative saved distance.

\begin{figure}
     \centering
     \begin{subfigure}[b]{0.49\textwidth}
         \includegraphics[width=\textwidth]{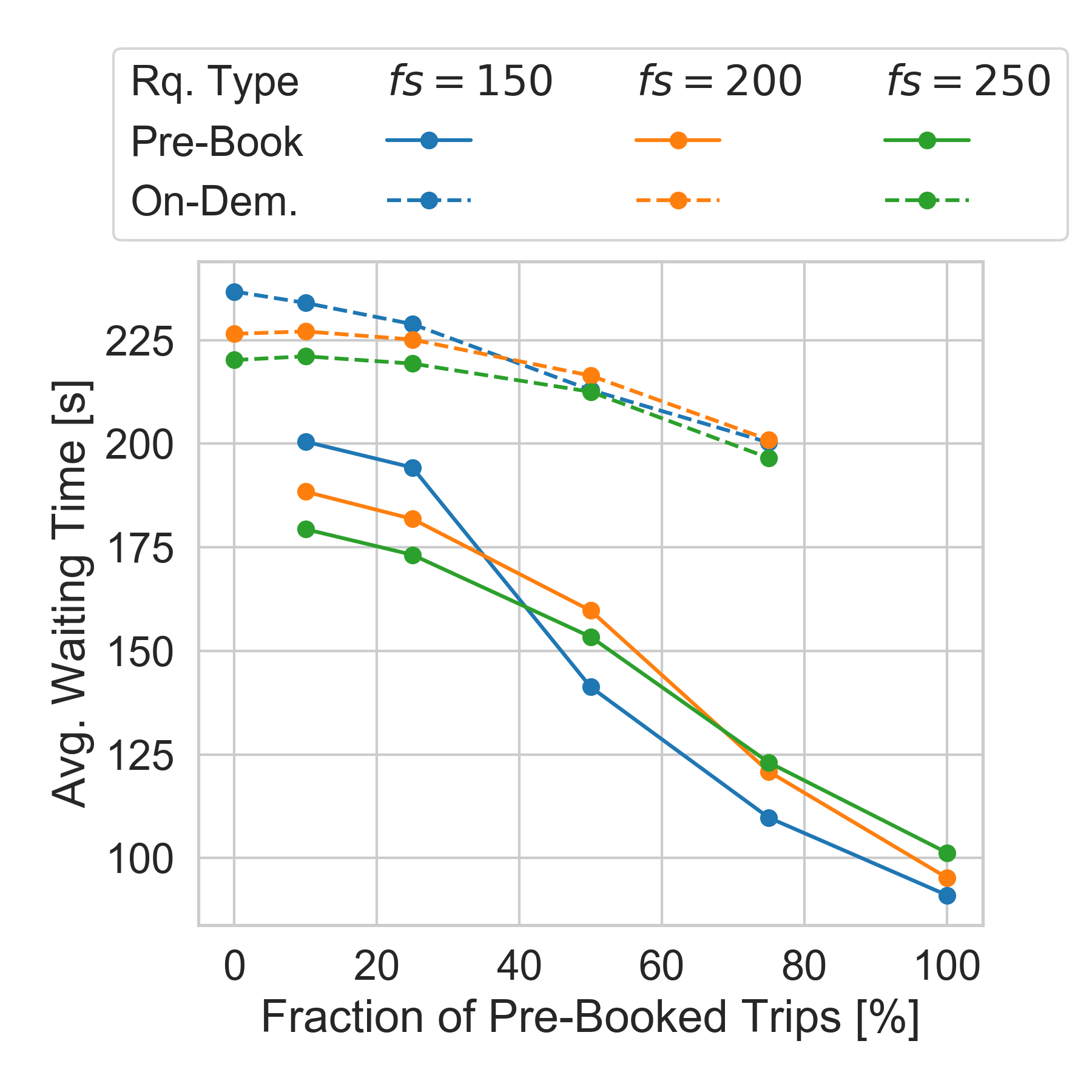}
         \caption{}
         \label{fig:waita}
    \end{subfigure}
     \hfill
     \begin{subfigure}[b]{0.49\textwidth}
         \includegraphics[width=\textwidth]{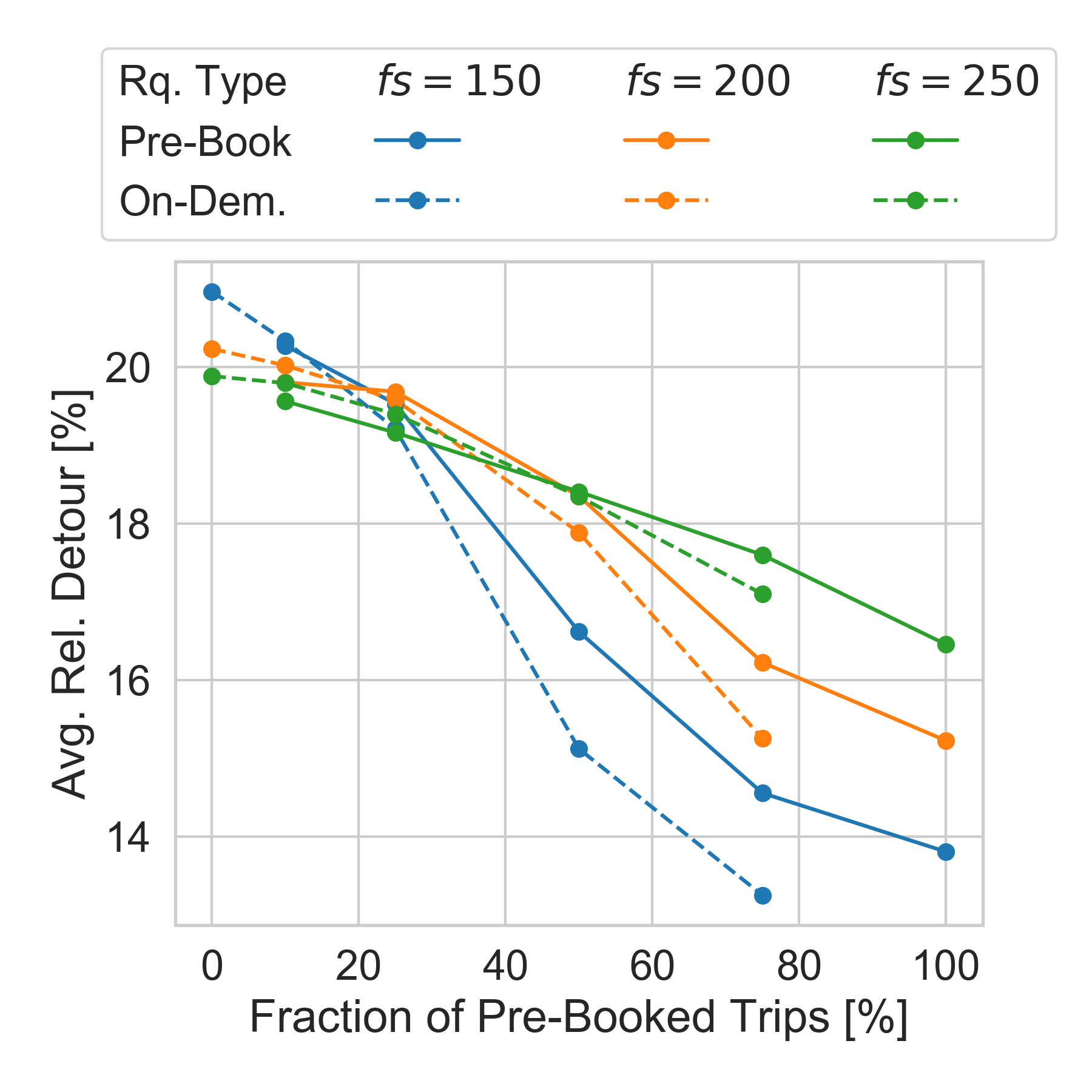}
         \caption{}
         \label{fig:detb}
    \end{subfigure}
        \caption{Waiting time and detour time for reservation and pre-booking customers. Only scenarios with repositioning are shown.}
        \label{fig:dem_cust}
\end{figure}

Finally, Fig.~\ref{fig:dem_cust} shows KPIs regarding the customer experience. Waiting and detour time are evaluated for both on-demand and pre-booking customers. For clarity, only scenarios with repositioning are illustrated. It can be observed that waiting times are reduced for pre-booking customers in comparison to on-demand customers because vehicles can wait at the pick-up position before their pre-booked pick-up time. Their detour is slightly increased because of more shared rides. Also on-demand customers benefit from pre-booking as for both, pre-booking and on-demand customers, waiting time and detour decrease with an increasing fraction of pre-booked trips.
     
\section{Conclusion and Future Work}

This study evaluated a mixed operation of on-demand and pre-booking customers for a ride-pooling service. Customers request a pre-booked ride by specifying an earliest pick-up time. The operator evaluates the request and replies whether this pre-booking can be fulfilled. Both, request as well as reply is considered to be carried out the day before the trip in this study. An optimization-based algorithm grouping requests in batches and connecting these batches to create schedules for pre-booked trips is proposed and compared to an insertion heuristic. Based on two rolling horizons, a method is proposed that dynamically infuses the solution for pre-booked customers into an online optimization algorithm, which assigns schedules to vehicles to serve on-demand and pre-booked customers together.

The method is tested for a case study of Manhattan, NYC. The proposed optimization algorithm outperforms the basic insertion heuristic for computing schedules for pre-booked customers: It provides better solutions up to ten times faster in the tested scenarios. When evaluating the impact of different penetrations of pre-booking customers, benefits for operator as well as customers can be observed: Due to additional knowledge major improvements in the optimization objective can be observed. Compared to a pure on-demand service more customers can be served while pooling efficiency can be maintained or even slightly increased. Nevertheless, for high penetrations of pre-booking customers the solution quality deteriorates, which can be traced back to the forward batch algorithm failing to accommodate all requests. While pre-booking customers benefit from an early booking confirmation and additionally decreased waiting times, also the waiting and detour times of on-demand customers decrease when pre-booking is enabled.

In future work the scaling property of the algorithm to compute the schedules for pre-booked customers, which induces rejections of pre-booking customers for high penetrations of pre-bookings, will be improved. This could be achieved by adding a meta-heuristic to improve the solution or reformulate the batching procedure in the forward batch optimization algorithm. For a real-world application some simplifications of the study have to be removed: Varying network travel times should be considered when creating the offline schedules to create schedules that can be fulfilled when the traffic state changes. Additionally, the forward batch optimization algorithm can be extended easily for customers to dynamically pre-book their trip and not only the day before. Finally, forecasts of on-demand customers can be incorporated when creating the offline schedules to provide even better solutions for the mixed service.

\section*{Author Contributions}
The authors confirm contribution to the paper as follows: study conception and design: Roman Engelhardt, Florian Dandl, Klaus Bogenberger; data collection: Roman Engelhardt; analysis and interpretation of results: Roman Engelhardt, Florian Dandl; draft manuscript preperation: Roman Engelhardt. All authors reviewed the results and approved the final version of the manuscript.

%Bibliography
\bibliographystyle{plainnat}
\bibliography{bib}

\end{document}